\documentclass[12pt]{article}
\usepackage[dvipsnames]{xcolor}
\usepackage{jheppub}
\usepackage{amsmath,graphicx,braket}
\usepackage{orcidlink}
\usepackage{tikz,tikz-cd}

\def\nn{ \nonumber \\ }
\def\rd{ {\rm d}}

\def\CO{ {C}}
\def\CR{ {D}}
\def\O{ {\mathcal O}}
\def\R{ {\mathcal R}}
\def\msbar{$\overline{\text{MS}}$}
\def\bphi{\bar \phi}
\def\hphi{\widehat \phi}
\def\tphi{\widetilde \phi}

\usepackage{soul}

\title{Field Redefinitions and Infinite Field Anomalous Dimensions
}

\author[a]{Aneesh V.~Manohar\orcidlink{0009-0004-5497-8554},}

\author[a]{Julie Pag\`es,}

\author[b,c]{Jasper Roosmale Nepveu\orcidlink{0000-0001-9001-6775}}

\affiliation[a]{Physics Department 0319,
University of California San Diego,\\ 9500 Gilman Drive, La Jolla, CA 92093-0319, USA}
\affiliation[b]{Humboldt-Universit\"at zu Berlin, Institut f\"ur Physik, Newtonstr.\ 15, 12489 Berlin, Germany}
\affiliation[c]{Deutsches Elektronen-Synchrotron DESY, Notkestr.\ 85, 22607 Hamburg, Germany}

\emailAdd{amanohar@ucsd.edu}
\emailAdd{jcpages@ucsd.edu}
\emailAdd{jasperrn@physik.hu-berlin.de}

\abstract{Field redefinitions are commonly used to reduce the number of operators in the Lagrangian by removing redundant operators and transforming to a minimal operator basis. We give a general argument  that such field redefinitions, while leaving the $S$-matrix invariant and consequently finite, lead not only to infinite Green's functions, but also to infinite field anomalous dimensions $\gamma_\phi$. These divergences cannot be removed by counterterms without reintroducing redundant operators.
}

\allowdisplaybreaks

\begin{document} 
\begin{flushright}
DESY-24-020
\\
HU-EP-24/04-RTG%
\vspace{-1.6cm}
\end{flushright}
\maketitle
\flushbottom

\section{Introduction}

The $S$-matrix of quantum field theories is unchanged by field redefinitions~\cite{Chisholm:1961tha,Politzer:1980me,Arzt:1993gz,Manohar:2018aog}, so that Lagrangians related by field redefinitions are equivalent, and give the same physical theory. While the $S$-matrix remains invariant under field redefinitions,  Green's functions can (and do) change. Field redefinitions are often used to reduce the number of operators in the Lagrangian, and their couplings, to a minimal basis. In general, working in a minimal basis unavoidably leads to Green's functions and field anomalous dimensions which are infinite even after the addition of renormalization counterterms, even though the $S$-matrix is finite. A classic example of this phenomenon occurs with the penguin diagrams in the low-energy theory of weak interactions (see the discussion in \cite[\S6]{Manohar:2018aog}), where Green's functions are infinite starting at one-loop order. Green's functions cannot be made finite by a simple rescaling of the fields --- any attempt to make them finite reintroduces the redundant operators which were eliminated to obtain a minimal basis. This observation is relevant for theories of inflation, where one computes fluctuations from correlation functions of quantum fields. We also show that field anomalous dimensions are in general infinite starting at two-loop order when redundant operators are removed by field redefinitions.

Start with an EFT Lagrangian including all allowed operators which contribute to the action, \textit{i.e.} all operators that are not total derivatives. This is equivalent to reducing the set of operators by using only integration-by-parts identities. The resulting set of operators is referred to as a Green's basis in the literature.  Only a subset of operators in the Green's basis is independent under field redefinitions. The choice of independent operators is arbitrary, but the number of them is not.  The independent operators (in some convention) are referred to as ``physical'' operators $\O_i$, and the remaining ones are referred to as ``redundant'' operators $\R_j$. The Green's basis has both sets of operators $\{\O,\R\}$. The Lagrangian coefficients of the physical operators are denoted by $\CO_i$ and of the redundant operators by $\CR_j$. Field redefinitions can remove the redundant operators from the Lagrangian and modify the coefficients from $\{\CO,\CR\} \to \{\overline \CO,0\}$.  The resulting Lagrangian and coefficients will be referred to as being in the physical basis.

The Lagrangian in the Green's basis is renormalized in the \msbar\  scheme. The Lagrangian has counterterms which depend on $\{\CO,\CR\}$, and Green's functions and $S$-matrices computed with the renormalized Lagrangian are finite. The $\beta$-functions and field anomalous dimension when $\epsilon \to 0$ are finite,
\begin{align}
\mu \frac{\rd \CO_i}{\rd \mu} &= \beta_{\CO_i}(\{\CO,\CR\}) \,, &
\mu \frac{\rd \CR_i}{\rd \mu} &= \beta_{\CR_i}(\{\CO,\CR\}) \,, &
\mu \frac{\rd  \phi}{\rd \mu} &=  -\gamma_\phi(\{\CO,\CR\})\,\phi \,,
\label{20}
\end{align}
and depend on all the parameters in the Lagrangian. After a transformation to the physical basis, the $\beta$-functions and field anomalous dimensions have the form
\begin{align}
\mu \frac{\rd \overline \CO_i}{\rd \mu} &= \beta_{\overline \CO_i}(\{\overline \CO\}) \,, &
\mu \frac{\rd  \phi}{\rd \mu} &=  -\gamma_\phi(\{\overline \CO\})\,\phi \,,
\label{21}
\end{align}
and depend only on the physical couplings. The $\beta$-functions are finite, but Green's functions and the field anomalous dimension $\gamma_\phi$ are \emph{infinite}, as was recently encountered in a specific case in ref.~\cite{Jenkins:2023bls}. 

Infinite Green's functions and field anomalous dimensions generically arise from field redefinitions. Start with the renormalized Lagrangian in the physical basis. Loop graphs computed with insertions of only the physical operators $\O$ can still lead to divergences which require counterterms with redundant operators $\R$. These divergences induce non-zero values for the redundant coefficients $\CR$ which are $1/\epsilon^k$ poles and generate $\beta$\nobreakdash-functions for the redundant couplings: $\mu \,\rd D_i/\rd\mu \neq 0$. These are, however, obscured because the theory is parametrized at the special point in theory space with  $\CR(\bar \mu)=0$. Nevertheless, the $\beta$-functions of the physical couplings and the field anomalous dimension depend on the counterterms of the redundant operators. An additional field redefinition is required to remove the counterterms of the redundant operators, and thereby transform the Lagrangian back to the physical basis, such that  $\overline \CR(\mu)=0$ for all $\mu$, and the bare coupling of redundant operators vanishes, $\overline \CR_b=0$. This field redefinition is infinite, since it removes counterterm coefficients of redundant operators. Since the $S$-matrix is invariant under field redefinitions, and remains finite, this means that any resulting $\beta_{\overline \CO_i}$ is finite.  However, Green's functions are modified by this field redefinition, and become infinite, typically starting at one-loop order. Likewise, the field anomalous dimension has $1/\epsilon$ poles starting at two-loop order when the scale dependence of the redundant couplings $\CR_i$ is ignored.

Infinite $\beta$-functions were found recently for the Standard Model Yukawa couplings at three-loop order~\cite{Bednyakov:2014pia,Herren:2017uxn,Herren:2021yur}. The origin of the divergence is due to an infinite $\mu$-dependent flavor rotation, and has a different origin than the divergences studied in this paper.

We now demonstrate the above results with an explicit computation in the $O(n)$ EFT to two-loop order.

\section{Example}

The example theory is the $O(n)$ EFT to dimension six with Lagrangian
\begin{align}
\mathcal{L} &= \frac12  (\partial_\mu  \phi_b \cdot \partial^\mu  \phi_b) - \frac12 m_b^2 (\phi_b \cdot \phi_b) - \frac14 \lambda_b
(\phi_b \cdot \phi_b)^2  +  \CO_{4,b}   \O_{4,b} 
 +  \CR_{4,b} \R_{4,b}  + \CO_{6,b} \O_{6,b} + \CR_{2,b} \R_{2,b}  \nn
&= \frac12 Z_\phi (\partial_\mu  \phi \cdot \partial^\mu  \phi) - \frac12  Z_\phi Z_{m^2} m^2 (\phi \cdot \phi) - \frac14 \mu^{2 \epsilon} Z_\phi^2  Z_\lambda  \lambda
 (\phi \cdot \phi)^2  \nn
&+ \mu^{2 \epsilon}  Z_\phi^2  Z_{\CO_4} \CO_4   \O_4  + \mu^{2 \epsilon} Z_\phi^2Z_{\CR_4}   \CR_4    \R_4  + \mu^{4 \epsilon} Z_\phi^3 Z_{\CO_6} \CO_6      \O_6 +  Z_\phi Z_{\CR_2} \CR_2 \R_2 \,,
\label{eq:Loff-shell}
\end{align}
where $\phi$ is an $n$-component real scalar field. The subscripts $b$ refer to bare quantities. The dimension six terms are
\begin{align}
\begin{aligned}
	\O_4&= (\partial_\mu  \phi \cdot \partial^\mu  \phi) (\phi \cdot \phi) \,,
	& \qquad  & \R_4= (\phi \cdot \partial_\mu \phi)^2\,, \\
	\O_6&= (\phi \cdot  \phi)^3 \,,
	& \qquad & \R_2= (\partial_\mu \partial^\mu  \phi \cdot \partial_\nu \partial^\nu  \phi )\,,
\end{aligned}
\label{6}
\end{align}
where we have divided the dimension-six operators into ``physical'' operators $\O_{4,6}$ and ``redundant'' operators $\R_{4,2}$. The subscript denotes the number of fields in the operator. The $O(n)$ EFT has an expansion in a mass scale $M$, so  the dimension-six coefficients $\CO_4, \CO_6, \CR_4, \CR_2$ are order $1/M^2$, and terms of higher order in $1/M$ are neglected in eq.~\eqref{eq:Loff-shell}. The physical operator coefficients are denoted collectively by $\{\CO\}$, and the redundant operator coefficients by $\{ \CR\}$. We include the dimension-two mass term $(\phi \cdot \phi)$ and dimension-four $(\phi \cdot \phi)^2$ interaction in the physical operators, and $m^2$ and $\lambda$ in the physical coefficients.

One can make a field redefinition in eq.~\eqref{eq:Loff-shell} to eliminate two of the dimension-six operators. Our choice in this paper is to eliminate $\R_{4,2}$ and retain $\O_{4,6}$, so that the minimal basis of dimension-six operators is $\{\O_4,\O_6\}$. The choice of minimal operator basis is arbitrary, but the number of minimal operators is the same in any basis. All dimension-six operators $\{\O_4,\O_6,\R_4,\R_2\}$ are included in the Green's basis.

The Lagrangian eq.~\eqref{eq:Loff-shell} in the Green's basis can be renormalized in dimensional regularization in the \msbar\ scheme. The counterterms to two-loop order and dimension-six are given in Appendix.~\ref{sec:gbct}, and the $\beta$-functions and field anomalous dimension are given in Appendix.~\ref{sec:gbanomdim}. The field anomalous dimension and $\beta$-functions are all finite, and  are functions of all the parameters in eq.~\eqref{eq:Loff-shell}. The 't~Hooft consistency conditions~\cite{tHooft:1973mfk} for the counterterms given in~\cite[\S 6]{Jenkins:2023rtg} are satisfied, which implies that the $\beta$-functions and field anomalous dimensions are finite.

The field redefinition
\begin{align}
\phi_b &\to \phi_b +  f\, \phi_b (\phi_b \cdot \phi_b)+ g\, \partial^2 \phi_b  \,, 
\label{13}
\end{align}
can be used to eliminate redundant operators in the Lagrangian.  $f$ and $g$ are functions of the bare couplings of order $1/M^2$, and independent of $\mu$, so the field-redefinition eq.~\eqref{13} preserves $\mu$-independence of the Lagrangian. The field redefinition
\begin{align}
 \phi_b &\to h \, \phi_b \,,
 \label{eq:phi_rescaling}
\end{align}
with $h$ a function of the bare couplings corresponds to a simple rescaling of the bare field and will modify $Z_\phi$ while keeping  $Z$ of the couplings unchanged. Equations~\eqref{13} and \eqref{eq:phi_rescaling} comprise the most general field redefinition compatible with $O(n)$ invariance to order $1/M^2$.

There are two independent terms in eq.~\eqref{13}, so we can eliminate at most two operators from the Lagrangian, which have been chosen to be $\R_4$ and $\R_2$. The general form for $f$ and $g$ must be compatible with the EFT power counting, so that the new Lagrangian retains the $1/M$ expansion. In addition, the dimensions of the terms must match in $4-2\epsilon$ dimensions, where  coupling constant dimensions can be fractional, e.g.\ $\lambda_b$ has dimension $2 \epsilon$. The allowed redefinition is
\begin{align}
\phi_b &=  \hphi_b +  (a_1 \CR_{4,b} + a_2 \lambda_b \CR_{2,b} )  \hphi_b ( \hphi_b \cdot  \hphi_b)+ a_3 \CR_{2,b}\, \partial^2  \hphi_b  \,,
\label{14}
\end{align}
in terms of bare fields, or
\begin{align}
\phi &=  \hphi  +  (a_1 Z_{\CR_4} \CR_{4} + a_2 Z_{\CR_2} Z_\lambda \lambda \CR_{2} ) Z_\phi \mu^{2\epsilon} \hphi ( \hphi \cdot  \hphi)+ a_3 Z_{\CR_2} \CR_{2}\, \partial^2  \hphi  \,,
\label{14a}
\end{align}
in terms of renormalized fields where $a_i$ are numbers, and the Lagrangian becomes
\begin{align}
\mathcal{L} &= \frac12 \left[1+2 a_3 m_b^2 \CR_{2,b} \right] (\partial \hphi_b \cdot \partial \hphi_b) - \frac12 m_b^2 (\hphi_b \cdot \hphi_b) \nn
&- \frac14
\left[ \lambda_b + 4 a_1 m_b^2  \CR_{4,b} +4  a_2 m_b^2 \lambda_b \CR_{2,b} )  \right] 
(\hphi_b \cdot \hphi_b)^2  \nn
& +  \left[ \CO_{4,b} + a_1 \CR_{4,b} + (a_2+a_3) \lambda_b \CR_{2,b} ) \right]  (\partial \hphi_b \cdot \partial \hphi_b)  (\hphi_b \cdot \hphi_b) \nn
& +  \left[ \CR_{4,b} + 2 a_1 \CR_{4,b} +2 (a_2+a_3) \lambda_b \CR_{2,b} )\right] (\hphi_b \cdot \partial \hphi_b)^2 \nn
&  + \left[\CO_{6,b}- a_1 \lambda_b \CR_{4,b} - a_2 \lambda_b^2 \CR_{2,b} ) \right]  (\hphi_b \cdot \hphi_b)^3 + 
\left[1-a_3\right] \CR_{2,b} (\partial^2 \hphi_b \cdot \partial^2 \hphi_b) \,.
\label{eq:L1}
\end{align}
In terms of renormalized couplings and fields, the Lagrangian is eq.~\eqref{eq:L1} with $\CO_b  \to Z_\CO \mu^{f_C \epsilon} \CO$, $\CR_b \to Z_\CR \mu^{f_D \epsilon} \CR$ and $\hphi_b \to \sqrt{Z_\phi}\, \hphi$, where $f_i = F_i -2$, and $F_i$ is the number of fields in $\O_i$, which determines the fractional part of the classical scaling dimension of the operator in $4-2\epsilon$ dimensions. The field renormalization for $\hphi$ is then $Z_{\hphi}=(1 + 2a_3 Z_{m^2} Z_{\CR_2} m^2 \CR_{2})Z_\phi$.

Note that the kinetic term is no longer canonically normalized, not even at tree level. The additional rescaling
\begin{align}
\hphi_b &=  \left[1- a_3 m_b^2 \CR_{2,b} \right]  \tphi_b \,, &
\hphi &=  \left[1 - a_3 Z_{\CR_2} Z_{m^2}m^2 \CR_{2} \right]  \tphi \,,
\label{15}
\end{align}
transforms the Lagrangian to
\begin{align}
\mathcal{L} &= \frac12 (\partial   \tphi_b \cdot \partial   \tphi_b) - \frac12 m_b^2  \left[1-2 a_3 m_b^2 \CR_{2,b} \right]  (  \tphi_b \cdot   \tphi_b) \nn
& - \frac14
\left[ \lambda_b + 4 a_1 m_b^2  \CR_{4,b} +4  (a_2-a_3) m_b^2 \lambda_b \CR_{2,b}  \right] 
(  \tphi_b \cdot   \tphi_b)^2  \nn
& +  \left[ \CO_{4,b} + a_1 \CR_{4,b} + (a_2+a_3) \lambda_b \CR_{2,b} \right]  (\partial   \tphi_b \cdot \partial   \tphi_b)  (  \tphi_b \cdot   \tphi_b) \nn
& +  \left[ \CR_{4,b} + 2 a_1 \CR_{4,b} +2 (a_2+a_3) \lambda_b \CR_{2,b} \right] (  \tphi_b \cdot \partial   \tphi_b)^2 \nn
&  + \left[\CO_{6,b}- a_1 \lambda_b \CR_{4,b} - a_2 \lambda_b^2 \CR_{2,b}  \right]  (  \tphi_b \cdot   \tphi_b)^3 + 
\left[1-a_3\right] \CR_{2,b} (\partial^2   \tphi_b \cdot \partial^2   \tphi_b) \,,
\label{eq:L2}
\end{align}
restoring canonical normalization of the kinetic energy term, and gives $Z_{\tilde \phi}= Z_\phi$.

Comparing with the original Lagrangian in eq.~\eqref{eq:Loff-shell} gives the transformed coefficients ($\widetilde \CO, \widetilde \CR$)
\begin{align}
\widetilde m_b^2 &= m_b^2  \left[1-2 a_3 m_b^2 \CR_{2,b} \right]   \,, \nn
\widetilde \lambda_b &=   \lambda_b + 4 a_1 m_b^2  \CR_{4,b} +4  (a_2-a_3) m_b^2 \lambda_b \CR_{2,b}  \,,   \nn
\widetilde \CO_{4,b} &=  \CO_{4,b} + a_1 \CR_{4,b} + (a_2+a_3) \lambda_b \CR_{2,b} \,,   \nn
\widetilde \CO_{6,b} &=   \CO_{6,b}- a_1 \lambda_b \CR_{4,b} - a_2 \lambda_b^2 \CR_{2,b} \,,  \nn
\widetilde \CR_{4,b} &=  \CR_{4,b} + 2 a_1 \CR_{4,b} +2 (a_2+a_3) \lambda_b \CR_{2,b} \,,   \nn
\widetilde \CR_{2,b} &= \left( 1-a_3\right) \CR_{2,b} \,,
\label{17}
\end{align}
which are functions of the original couplings and $a_i$. The choice $ a_1=-1/2$, $ a_2=-1$, $ a_3=1$ gives $\widetilde \CR_{4,b} =0$, $\widetilde \CR_{2,b}=0$, so that the redundant operators are eliminated. The new bare couplings in the physical basis $(\overline \CO, \overline \CR)$ are functions of the original bare couplings,
\begin{align}
\overline m_b^2 &= m_b^2  \left[1-2  m_b^2 \CR_{2,b} \right]   \,, &
\bar{\lambda}_b &=   \lambda_b -2 m_b^2  \CR_{4,b} - 8 m_b^2 \lambda_b \CR_{2,b}  \,,   \nn
\overline \CO_{4,b} &=  \CO_{4,b} -\frac12 \CR_{4,b}  \,,   &
\overline \CO_{6,b} &=   \CO_{6,b} + \frac12 \lambda_b \CR_{4,b} +  \lambda_b^2 \CR_{2,b} \,,  \nn
\overline \CR_{4,b} &=  0 \,,   &
\overline \CR_{2,b} &= 0 \,.
\label{22}
\end{align}
and are the values of  ($\widetilde \CO, \widetilde \CR$) at $ a_1=-1/2$, $ a_2=-1$, $ a_3=1$.

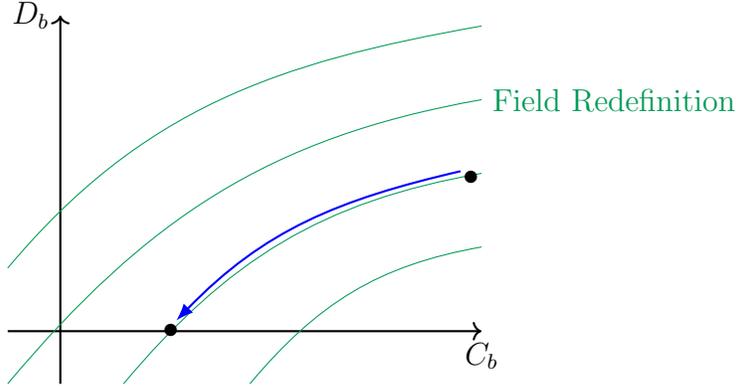
\begin{figure}
\begin{center}
\begin{tikzpicture}[scale=1.4,
mid/.style = {decoration={
    markings,mark=at position 0.65 with {\arrow{Latex}}}}]] 
			\draw[->, thick] (0,-0.5) -- (0,3) node[left] {$\CR_b$} ;
			\draw[->, thick] (-0.5,0) -- (4,0) node[below] {$\CO_b$} ;
 	
 	 \draw[color=ForestGreen]    (.6,-.5)  to[out=50,in=-170] (4,1.5);
 	  \draw[color=ForestGreen,yshift=.7cm]    (-0.5,-1.2)  to[out=50,in=-170] (4,1.5)node[right] { \textcolor{ForestGreen}{Field Redefinition}};
 	  \draw[color=ForestGreen,yshift=1.4cm]    (-0.5,-.8)  to[out=50,in=-170] (4,1.5);
 	  \draw[color=ForestGreen,yshift=-0.7cm]    (1.8,.2)  to[out=50,in=-170] (4,1.5);
 	  
 	  \draw[Latex-,color=blue,thick]    (1.1,0.1)  to[out=46,in=-167] (3.8,1.52);

 	  \node at (1.05,0){$\bullet$};
 	  \node at (3.9,1.46){$\bullet$};
 
\end{tikzpicture}
\end{center}
\caption{\label{fig:1b} Field redefinitions lead to a set of equivalent theories with the same (finite) $S$-matrix, shown by the green curves in the space of bare couplings. The coefficients $\widetilde \CO_b$ and $\widetilde \CR_b$ vary along the field redefinition curves as the parameters $\{a\}$ in the field redefinition are varied. The coefficients $\overline \CO_b$ are the values of $\widetilde \CO_b$ when the green curve intersects the $\CO_b$-axis and the redundant couplings $\CR_b$ vanish. The bare couplings are infinite.}
\end{figure}

In general, we have coefficients
\begin{align}
\widetilde  \CO_{i,b} &= F_i(\{a\},\{\CO_{b}\},\{ \CR_b \} )\,, &
\widetilde  \CR_{i,b} &= G_i(\{a\},\{\CO_{b}\},\{ \CR_b \} )\,,
\label{23}
\end{align}
which are functions of the field redefinition parameters $\{a\}$ and the original coefficients $\{\CO_b\},\{\CR_b\}$. To go to the physical basis, the parameters $\{a\}$ are chosen to set $\widetilde \CR_{i,b} =0$ giving
\begin{align}
\overline  \CO_{i,b} &= F_i(\{\CO_{b}\},\{ \CR_b \} )\,, &
\overline \CR_{i,b} &= 0 \,,
\label{24}
\end{align}
substituting the values for the parameters which set $\widetilde \CR =0$ back in eq.~\eqref{23}. The transformation is shown schematically in figure~\ref{fig:1b}.
%
%
%
%
%

The original theory was parameterized by $\CO$ and $\CR$. We can use eq.~\eqref{24} to determine $\overline \CO$, and parameterize the theory instead by $\overline \CO$ and $\CR$. We still need to retain $\CR$ so that eq.~\eqref{24} can be inverted to obtain the original couplings $\CO$ and $\CR$ from $\overline \CO$ and $\CR$. However, we emphasize that it is arbitrary to keep $\overline\CO$ and $\CR$. For example, in the $O(n)$ case we consider, it also valid to parametrize the theory by $\overline\CO$ and $\CO$. In general, one has to retain as many total parameters as in the original theory.

The new renormalized couplings are given by the same functions $F_i$ of the original renormalized couplings
\begin{align}
\overline  \CO_{i}(\mu) &= F_i(\{\CO(\mu)\},\{ \CR(\mu) \} )\,, &
\overline \CR_{i}(\mu) &= 0 \,,
\label{eq:RenCRedef}
\end{align}
obtained by dropping the $1/\epsilon$ terms in eq.~\eqref{24}. In the $O(n)$ example, the relations are
\begin{align}
\overline m^2(\mu) &= m^2(\mu)  \left[1-2  m^2(\mu) \CR_{2}(\mu) \right]   \,, \nn
\bar{\lambda}(\mu) &=   \lambda(\mu) -2 m^2(\mu)  \CR_{4}(\mu) - 8 m^2(\mu) \lambda(\mu) \CR_{2}(\mu)  \,,   \nn
\overline \CO_{4}(\mu) &=  \CO_{4}(\mu) -\frac12  \CR_{4}(\mu)  \,,   \nn
\overline \CO_{6,b} &=   \CO_{6,b} + \frac12 \lambda(\mu)   \CR_{4}(\mu) +   \lambda^2(\mu)  \CR_{2}(\mu)  \,,  \nn
\overline \CR_{4}(\mu)  &=  0 \,,   \nn
\overline \CR_{2}(\mu)  &= 0 \,.
\label{27}
\end{align}
The  renormalization for the field in eq.~\eqref{eq:L2} at $ a_1=-1/2$, $ a_2=-1$, $ a_3=1$, denoted by $\bphi$ is
\begin{align}
\bphi_b &= \sqrt{Z_{ \bphi}}\  \bphi & Z_{ \bphi} &= Z_\phi\,.
\label{27a}
\end{align}

The overall field redefinition that has been performed is a combination of eq.~\eqref{14} and eq.~\eqref{15}, and is infinite, leading to infinite Green's functions. The only way to restore finite Green's functions is to undo the field redefinition and reintroduce the redundant operators. A simple rescaling of $\phi$ does not make the Green's functions finite, since the transformation eq.~\eqref{14} is non-linear, and cannot be compensated for by a rescaling.

The new bare and
and renormalized couplings are related by
\begin{align}
\overline  C_{i,b} \, \mu^{-f_i \epsilon} &= \overline   C_i(\mu) + \overline   C_{i,\text{c.t.}}(\mu) =   Z_{\overline C_i} \overline  C_i(\mu) \,,
\label{26}
\end{align}
where $f_i $ is defined below eq.~\eqref{eq:L1}. The counterterms for $\overline \CO_i$ to two-loop order are given in Appendix.~\ref{sec:physct}, and the $\beta$-functions and field anomalous dimension are given in Appendix.~\ref{sec:physanomdim}. The $\beta$-functions for $\overline C_i$ depend only on the physical couplings $\overline m^2$, $\bar{\lambda}$, $\overline \CO_4$ and $\overline \CO_6$, and are finite. The field renormalization $Z_{\bphi}$ is a function of the physical couplings $\overline \CO$ as well as the original redundant couplings $\CR$ before the field redefinition.
The field anomalous dimension $\gamma_{\bphi}$ computed from the logarithmic derivative of $Z_{\bphi}$,
\begin{align}
	\gamma_{\bphi} \equiv \frac12 Z_{\bphi}^{-1}  \dot  Z_{\bphi} = \frac12 Z_{\bphi}^{-1} \left(  \sum_i \frac{\partial Z_{\bphi}}{\partial \overline \CO_i} \dot{\overline{\CO}}_i  + \sum_j \frac{\partial Z_{\bphi}}{\partial \CR_j} \dot \CR_j  \right) \,,
	\label{eq:gammaphidef}
\end{align}
where $\dot{C} \equiv \mu\,  \rd C / \rd \mu $, is
given in eq.~\eqref{eq:finitegammaphi}. It is finite, provided one includes both the $\overline \CO$ and $\CR$ terms in eq.~\eqref{eq:gammaphidef}.
Setting the $\{D\}$ to zero in $Z_{\bphi}$ before taking the derivative w.r.t to $D$ in \eqref{eq:gammaphidef} leads to a violation of the field anomalous dimension consistency condition given in~\cite[\S 6]{Jenkins:2023rtg} and generates an infinite piece for $\gamma_{\bphi}$ at two loop,
\begin{equation}
	\frac1{2\epsilon} \sum_i \frac{\partial \gamma_{\bphi}^{(0)} }{\partial D_i}\beta_{D_i}
\label{2.19}
\end{equation}
where $\gamma_{\bphi}^{(0)}$ is the field anomalous dimension at one loop, which is finite.

In order to remove the dependence of $Z_{\bphi}$ on $\CR$, we can perform an additional rescaling of the field
\begin{align} \label{eq:infinitefieldredef}
\bphi &=  \left[1+ a_4 \CR_{2} + a_5  \CR_{4} \right]  \check{\phi} \,,
\end{align}
giving $Z_{\check \phi}= Z_{\bphi} \left[1+ 2 a_4 \CR_{2} + 2 a_5  \CR_{4} \right]$,
with\footnote{The notation $\{\}_{1,2}$ denote the one and two-loop terms, and must be multiplied by $1/(16 \pi^2)$ and $1/(16 \pi^2)^2$, respectively. }
\begin{align}
	a_4 &= (n+2) \bar{\lambda}^2 \overline m^2 \, \left\{ \frac{1}{\epsilon} \right\}_2 \nn
	a_5 &= -\frac{1}{2} (n+2) \overline m^2 \left\{ \frac{1}{\epsilon} \right\}_1 +   \frac{7}{4} (n+2) \overline m^2 \bar{\lambda} \left\{ \frac{1}{\epsilon} \right\}_2
	-\frac{1}{2} (n+2) (n+5)  \overline m^2 \bar{\lambda} \left\{ \frac{1}{\epsilon^2} \right\}_2
\end{align}
which completely removes all $D_2$ and $D_4$ dependence from the Lagrangian. This shows that when working in the physical basis, \textit{i.e.} setting $D_2(\mu)=0$ and $D_4(\mu)=0$, we are implicitly making the infinite field redefinition eq.~\eqref{eq:infinitefieldredef} and using the field $\check{\phi}$. The anomalous dimension of $\check{\phi}$ given in eq.~\eqref{b.no} is infinite, because the logarithmic derivative of $1+ a_4 \CR_{2} + a_5  \CR_{4}$ does not vanish even at $\CR_{2,4}=0$ since the $\beta$-functions of $\CR_{2,4}$ do not vanish at that point, and $a_4$ and $a_5$ are infinite.

As noted before, it is equally valid to parametrize the theory by, for example, $\{\overline\CO_6, \overline\CO_4, \CO_4, \CR_2\}$ instead of  $\{\overline\CO_6, \overline\CO_4, \CR_4, \CR_2\}$. This affects only the form of $Z_{\bphi}$, leaving $Z$ of the physical couplings unchanged. Moreover, the field anomalous dimension (both the finite and infinite part) depends on the calculation, and not only on the final choice of physical basis. 
For example in the $O(n)$ model, one can use the physical basis $B_1=\{\O_4,\O_6\}$ or $B_2=\{\R_4,\O_6\}$. The field anomalous dimension computed in basis $B_2$ is different from that computed in basis $B_1$ and then converted to basis $B_2$ by an additional field redefinition for both the finite and infinite pieces. However the $S$-matrix computed in the two methods agrees.

Given that the field anomalous dimension is infinite starting at two-loop order, one can abandon attempts to renormalize the field, and simply use the bare field to compute $S$-matrix elements. In this case $\gamma_\phi=0$. However, now the two point function $\left\langle \phi(x) \phi(y) \right\rangle$ is infinite, and the (infinite) wavefunction factor has to be included in the $S$-matrix computation to obtain a finite $S$-matrix, which makes the computation more involved.  

%
%
\begin{figure}
\begin{center}
\begin{tikzpicture}[scale=1.4,
mid/.style = {decoration={
    markings,mark=at position 0.65 with {\arrow{Latex}}}}]] 
			\draw[->, thick] (0,-0.5) -- (0,3) node[left] {$D$} ;
			\draw[->, thick] (-0.5,0) -- (4,0) node[below] {$C$} ;

 	\draw[color=Red,, xshift=-1cm]    (0.5,0)  to[out=75,in=-145] (2.2,3);
  	\draw[color=Red]    (0.2,-.5)  to[out=80,in=-145] (2.2,3)node[above] { \textcolor{Red}{RG flow}};
   	\draw[color=Red, xshift=1cm]    (0.2,-.5)  to[out=80,in=-145] (2.2,3);
 	\draw[color=Red, xshift=2cm]    (0.2,-.5)  to[out=80,in=-145] (2,2.8);
 	\draw[color=Red, xshift=3cm]    (0.2,-.5)  to[out=80,in=-130] (1,1.6) ;
 	
 	\draw[-Latex,color=orange, xshift=2cm,thick]    (0.27,0.1)  to[out=80,in=-115] (.54,1.02);
 	
 	 \draw[color=ForestGreen]    (.6,-.5)  to[out=50,in=-170] (4,1.5);
 	  \draw[color=ForestGreen,yshift=.7cm]    (-0.5,-1.2)  to[out=50,in=-170] (4,1.5)node[right] { \textcolor{ForestGreen}{Field Redefinition}};
 	  \draw[color=ForestGreen,yshift=1.4cm]    (-0.5,-.8)  to[out=50,in=-170] (4,1.5);
 	  \draw[color=ForestGreen,yshift=-0.7cm]    (1.8,.2)  to[out=50,in=-170] (4,1.5);
 	  
  	  \draw[Latex-,color=blue,thick]    (1.17,0.03)  to[out=46,in=-153] (2.55,1.01);
 
  	\draw (2.62,1.10) circle (0.06);  
 	  \node at (1.05,0){$\bullet$};
 	  \node at (2.3,0){$\bullet$};
 	  \node[color=ForestGreen] at (4.25,1.5){$\mu_2$};
 	   \node[color=ForestGreen] at (4.25,0.8){$\mu_1$};

\end{tikzpicture}
\end{center}
\caption{\label{fig:1} Field redefinitions and RG flow lead to a set of equivalent theories with the same (finite) $S$-matrix, shown by the green and red curves respectively in the space of renormalized couplings. The coefficients $\widetilde \CO$ and $\widetilde \CR$ vary along the field redefinition curves as the parameters $\{a\}$ in the field redefinition are varied.  The coefficients $\overline \CO$ are the values of $\widetilde \CO$ when the green curve intersects the $\CO$-axis. The renormalized couplings are finite. Starting from vanishing redundant couplings $\CR(\mu_1)=0$ can still lead to non-zero redundant couplings $\CR(\mu_2)\not=0$ at a different value of $\mu$.}
\end{figure}
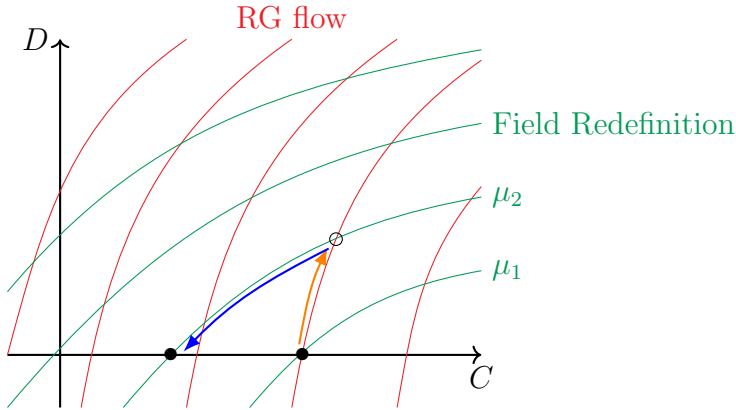
Figure~\ref{fig:1} shows the renormalized coupling constant space for the theory. (Figure~\ref{fig:1b} showed the bare coupling constant space.) In the space of renormalized couplings, we have two different flows. There is a flow due to field redefinitions, analogous to that in figure~\ref{fig:1b}. Along these flow lines, the $S$-matrix is invariant. In addition, we have a flow  due to a change in $\mu$ which also leaves the $S$-matrix invariant.\footnote{The renormalized couplings change, but the $S$-matrix remains invariant because the coupling constant dependence is canceled by $\log \mu^2/s$ terms in the formul\ae\ for $S$-matrix elements, where $s$ is a kinematic invariant with mass-squared dimension.} Since the $S$-matrix is invariant under both flows,  the two are compatible. A point on the field-redefinition curve at $\mu=\mu_1$ flows to some point on the field redefinition curve at $\mu=\mu_2$.

Suppose we compute loop corrections in the Green's basis starting with a renormalized Lagrangian with $\CR(\mu)=0$ at $\mu=\mu_1$. Even though the renormalized coupling vanishes, $\CR(\mu_1)=0$, loop corrections can generate counterterms for $\CR(\mu)$ which depend on the non-zero physical couplings $\CO(\mu)$. Thus RG evolution induces non-zero couplings $\CR(\mu)$ as $\mu$ evolves from $\mu_1$ to $\mu_2$. This flow is shown by the orange arrow in figure~\ref{fig:1}. One then needs to do a field redefinition at $\mu_2$ to make $\CR(\mu_2)=0$, shown by the blue arrow in figure~\ref{fig:1}. RG evolution in  the EFT with only physical couplings is equivalent to a combination of RG evolution and field redefinitions in the Green's basis. Performing the field redefinition to all orders in the counterterms is the same as the transformation using bare couplings discussed earlier. During all the transformations, the $S$-matrix is invariant, and remains finite. The $S$-matrix is determined by the physical couplings (and vice-versa), so they remain finite as well, and the RG flow for the physical couplings is finite. The evolution of the physical couplings (the black dot in figure~\ref{fig:1}) is determined by the intersection of the field-redefinition invariance curve with the $\CO$ axis; it does not depend on the starting point on the curve, \textit{i.e.} the field-redefinition curves flow to other field-redefinition curves under a change in $\mu$. As a result, the $\beta$-functions $\overline C(\mu)$ are finite, and only depend on $\overline C(\mu)$. The key point is that they do not depend on $\CR$.

In the above analysis, we have made use of (a) the invariance of the $S$-matrix under field redefinitions and (b) a one-to-one relation between the physical couplings and the $S$-matrix. These conditions do not apply to the quantum field $\phi$, Green's functions are not invariant under field redefinitions, and the field anomalous dimension is generally infinite. The infinity arises due to the additional rescaling eq.~\eqref{eq:infinitefieldredef} to remove $\CR$ dependence in $Z_\phi$, or equivalently, dropping the $\CR$ derivative in eq.~\eqref{eq:gammaphidef}.

If one wants to keep Green's functions and field anomalous dimensions finite, redundant operators cannot be ignored and one has to use the full Green's basis at all steps in the computations.

\subsection*{Acknowledgments}

We thank Xiaochuan Lu for helpful discussions. JRN thanks Franz Herzog for collaboration on related topics and for providing access to his implementation of the $R^*$ operation \cite{Herzog:2017bjx, Cao:2021cdt} to calculate the counterterms. 
This work is supported in part by the U.S.\ Department of Energy (DOE) under award numbers~DE-SC0009919. JRN is supported by the Deutsche Forschungsgemeinschaft (DFG, German Research Foundation) --- Projektnummer 417533893/GRK2575 ``Rethinking Quantum Field Theory''. This project has received funding from the European Union’s Horizon Europe research and innovation programme under the Marie Skłodowska-Curie Staff Exchange  grant agreement No 101086085 – ASYMMETRY.

\begin{appendix}

\section{Green's basis results}\label{sec:gb}

The counterterms, $\beta$-functions and field anomalous dimension of the $O(n)$ theory to two-loop order in the Green's basis of eq.~\eqref{eq:Loff-shell} are listed below, where
\begin{equation}
\mu \frac{\rd C_i}{\rd \mu } \equiv - \epsilon f_i C_i + \beta_{C_i}
\end{equation}
in $4 -2 \epsilon$ dimensions, with $f_i$ defined below eq.~\eqref{eq:L1}, 
and
\begin{equation}
\mu \frac{\rd  \phi}{\rd \mu} =  -\gamma_\phi\,\phi\,.
\end{equation}

\subsection{Counterterms}\label{sec:gbct}

The renormalization factors for the Lagrangian eq.~\eqref{eq:Loff-shell} are:
\begin{align}
Z_\phi &=1 +
\begin{bmatrix} 1 \\ C_4 \\ C_6 \\ D_4 \\ D_2
\end{bmatrix}^\intercal
 \begin{bmatrix}
0 & -\frac12 (n+2) \lambda^2  & 0  \\
2  m^2 n  & -(n+2)  \lambda m^2  &  2(n+1)(n+2) \lambda m^2 \\
0 & 0 & 0 \\ 
2 m^2 & -(n+2)  \lambda m^2 & 4  (n+2)\lambda m^2\\
0 & 6  (n+2)\lambda^2 m^2 & 0
\end{bmatrix}
 \begin{bmatrix} \left\{ \frac{1}{\epsilon} \right\}_1 \\ \left\{ \frac{1}{\epsilon} \right\}_2 \\ \left\{ \frac{1}{\epsilon^2} \right\}_2 \end{bmatrix} 
\end{align}
\begin{align}
Z_{m^2} &=1 +
\begin{bmatrix} 1 \\ C_4 \\ C_6 \\ D_4 \\ D_2
\end{bmatrix}^\intercal
\begin{bmatrix}
(n+2) \lambda   & -\frac{5}{2}  (n+2)\lambda ^2  & (n+2)(n+5) \lambda ^2  \\
 -4 n m^2  & 7(n+2) \lambda  m^2  & -2 (n+2)(7n+6)\lambda  m^2   \\
 0 & 0 & -6   (n+2)(n+4) m^2\\
 -4 m^2 & 7  (n+2)\lambda  m^2 & -26 (n+2)  \lambda  m^2\\
 -6 (n+2)  \lambda  m^2& 36  (n+2)\lambda ^2 m^2 & -18 (n+2)(n+5)  \lambda ^2 m^2 
\end{bmatrix}
 \begin{bmatrix} \left\{ \frac{1}{\epsilon} \right\}_1 \\ \left\{ \frac{1}{\epsilon} \right\}_2 \\ \left\{ \frac{1}{\epsilon^2} \right\}_2 \end{bmatrix} 
\end{align}
\begin{align}
Z_{\lambda}\lambda &= \lambda +
\begin{bmatrix} 1 \\ C_4 \\ C_6 \\ D_4 \\ D_2
\end{bmatrix}^\intercal
\begin{bmatrix}
(n+8) \lambda ^2  & -3(3 n+14)\lambda ^3  & (n+8)^2\lambda ^3  \\
 -4  (3 n+4) \lambda  m^2& 2(29 n+154)  \lambda ^2 m^2 & -12(n+3)(3n+14) \lambda ^2 m^2  \\
 -12(n+4)  m^2 & 144 (n+4)\lambda  m^2  & -36(n+4)(n+10) \lambda  m^2  \\
 -28 \lambda  m^2 & 2 (37 n+146) \lambda ^2 m^2 & -48 (3 n+14)\lambda ^2 m^2  \\
 -12 (n+8) \lambda ^2 m^2 & 4 (67 n+302)\lambda ^3 m^2  & -12 (3 n^2+52 n+188)\lambda ^3 m^2 
\end{bmatrix}
 \begin{bmatrix} \left\{ \frac{1}{\epsilon} \right\}_1 \\ \left\{ \frac{1}{\epsilon} \right\}_2 \\ \left\{ \frac{1}{\epsilon^2} \right\}_2 \end{bmatrix} 
\end{align}
\begin{align}
Z_{C_4}C_4 &= C_4 +
\begin{bmatrix} 1 \\ C_4 \\ C_6 \\ D_4 \\ D_2
\end{bmatrix}^\intercal
\begin{bmatrix}
 0 & 0 & 0 \\
 (n+6) \lambda  & -(9 n+16)\lambda ^2  &  \left(n^2+12 n+44\right)\lambda ^2 \\
 0 & 6   (n+4) \lambda& 0 \\
 \lambda  & \frac{1}{2} (5 n-4)\lambda ^2  & 2 (n+5)\lambda ^2  \\
 0 & (5 n+22)\lambda ^3  & 0
\end{bmatrix}
 \begin{bmatrix} \left\{ \frac{1}{\epsilon} \right\}_1 \\ \left\{ \frac{1}{\epsilon} \right\}_2 \\ \left\{ \frac{1}{\epsilon^2} \right\}_2 \end{bmatrix} 
\end{align}
\begin{align}
Z_{C_6}C_6 &= C_6 +
\begin{bmatrix} 1 \\ C_4 \\ C_6 \\ D_4 \\ D_2
\end{bmatrix}^\intercal
\begin{bmatrix}
 0 & 0 & 0 \\
 (n+8)\lambda ^2  & -3  (5 n+58)\lambda ^3 &\left(3 n^2+47 n+274\right)  \lambda ^3 \\
 3 (n+14)\lambda   & -\frac{3}{2} (53 n+394) \lambda ^2 & 3 (n+14)(2n+25)\lambda ^2  \\
 9 \lambda ^2 & - (23 n+166)\lambda ^3 & 4 (8 n+73)\lambda ^3  \\
  (n+26)\lambda ^3 & -(61 n+506)\lambda ^4  & 3 (n+11)(n+26)\lambda ^4 
\end{bmatrix}
 \begin{bmatrix} \left\{ \frac{1}{\epsilon} \right\}_1 \\ \left\{ \frac{1}{\epsilon} \right\}_2 \\ \left\{ \frac{1}{\epsilon^2} \right\}_2 \end{bmatrix} 
\end{align}
\begin{align}
Z_{D_4}D_4 &= D_4 +
\begin{bmatrix} 1 \\ C_4 \\ C_6 \\ D_4 \\ D_2
\end{bmatrix}^\intercal
\begin{bmatrix}
 0 & 0 & 0 \\
 -2(n-2) \lambda   & -(n-2)\lambda ^2  & -4  (n-2) (n+5)\lambda ^2 \\
 0 & 12(n+4)  \lambda  & 0 \\
 2  (n+3)\lambda  & -\frac{7}{2} (n+6) \lambda ^2 & \left(3 n^2+22 n+44\right)\lambda ^2  \\
 0 & 2(5 n+22) \lambda ^3  & 0 
 \end{bmatrix}
 \begin{bmatrix} \left\{ \frac{1}{\epsilon} \right\}_1 \\ \left\{ \frac{1}{\epsilon} \right\}_2 \\ \left\{ \frac{1}{\epsilon^2} \right\}_2 \end{bmatrix} 
\end{align}
\begin{align}
Z_{D_2}D_2 &= D_2 +
\begin{bmatrix} 1 \\ C_4 \\ C_6 \\ D_4 \\ D_2
\end{bmatrix}^\intercal
\begin{bmatrix}
 0 & 0 & 0 \\
 0 & \frac{1}{6} (n+2)\lambda   & 0 \\
 0 & 0 & 0 \\
 0 & \frac{1}{6}  (n+2) \lambda & 0 \\
 0 & \frac{1}{2}  (n+2)\lambda ^2 & 0 
 \end{bmatrix}
 \begin{bmatrix} \left\{ \frac{1}{\epsilon} \right\}_1 \\ \left\{ \frac{1}{\epsilon} \right\}_2 \\ \left\{ \frac{1}{\epsilon^2} \right\}_2 \end{bmatrix} 
\end{align}
\\

\subsection{\texorpdfstring{$\beta$}{beta}-Functions and \texorpdfstring{$\gamma_\phi$}{gamma\_phi}}\label{sec:gbanomdim}

The $\beta$-functions and field anomalous dimensions computed from the counterterms are
\begin{align} \nonumber
	\beta_{m^2} = ~&
\Big\{2 (n+2)\lambda  m^2  -8 n m^4  C_4-8  m^4 D_4-12 (n+2) \lambda  m^4 D_2 
\Big\}_1
\\\nonumber
&+ \Big\{ -10 (n+2) \lambda ^2 m^2  + 28 (n+2)  \lambda  m^4 C_4 +28 (n+2) \lambda  m^4 D_4+144 (n+2)  \lambda ^2 m^4 D_2
\Big\}_2 
\\\nonumber
	\beta_{\lambda} = ~&
\Big\{ 2 (n+8) \lambda ^2   -8 (3 n+4)  \lambda  m^2 C_4-24 (n+4)  m^2 C_6 -56  \lambda  m^2 D_4
\\\nonumber
&\quad -24 (n+8)  \lambda ^2 m^2 D_2
\Big\}_1
\\\nonumber
&+ \Big\{ -12 (3 n+14) \lambda ^3  + 8 (29 n+154)  \lambda ^2 m^2 C_4 + 576 (n+4)  \lambda  m^2 C_6 
\\\nonumber
&\quad +8 (37 n+146)  \lambda ^2 m^2 D_4+16 (67 n+302)  \lambda ^3 m^2 D_2 
\Big\}_2 
\\\nonumber
	\beta_{C_4} = ~&
\Big\{ 2 (n+6)  \lambda C_4 +2  \lambda D_4
 \Big\}_1
 \\\nonumber 
&+ \Big\{ -4 (9 n+16)  \lambda ^2 C_4 +24 (n+4)  \lambda C_6 +2 (5 n-4)  \lambda ^2 D_4+4 (5 n+22) \lambda ^3  D_2
\Big\}_2 
\\\nonumber
	\beta_{C_6} = ~&
\Big\{  2 (n+8)  \lambda ^2 C_4+ 6 (n+14)  \lambda C_6 +18  \lambda ^2 D_4+2 (n+26)  \lambda ^3 D_2
 \Big\}_1
 \\\nonumber
&- \Big\{12 (5 n+58)  \lambda ^3 C_4 +6 (53 n+394)  \lambda ^2 C_6 +4 (23 n+166) \lambda ^3  D_4+4 (61 n+506)  \lambda ^4 D_2
\Big\}_2 
\\\nonumber
	\beta_{D_4} = ~&
\Big\{-4 (n-2) \lambda  C_4  + 4 (n+3) \lambda  D_4   
 \Big\}_1
 \\\nonumber
&+ \Big\{  -4 (n-2) \lambda ^2 C_4  +48 (n+4) \lambda C_6 -14  (n+6) \lambda ^2 D_4  +8 (5 n+22) \lambda ^3 D_2  
\Big\}_2 
\\
	\beta_{D_2} = ~&
 \Big\{ \frac{2}{3} (n+2)  \lambda C_4  +\frac{2}{3}(n+2)  \lambda D_4 +2 (n+2)  \lambda^2 D_2     
  \Big\}_2 
\\ \nonumber
	\gamma_\phi = ~&
\Big\{ -2  n m^2  C_4-2 m^2  D_4
\Big\}_1\\ 
&+ \Big\{ (n+2) \lambda ^2 + 2 (n+2) \lambda  m^2 C_4 +2 (n+2) \lambda  m^2 D_4 -12 (n+2)  \lambda ^2 m^2 D_2
\Big\}_2
\end{align}

\section{Physical basis results}\label{sec:phys}

The counterterms, $\beta$-functions, and field anomalous dimension in the physical basis with physical couplings $\overline C$ and  redundant couplings $\overline D=0$ are listed below.  Note that the wavefunction renormalization $Z_{\bphi}$ depends on the redundant couplings $\CR$, which parametrized the Lagrangian \emph{before} the field redefinition.

\subsection{Counterterms}\label{sec:physct}

The renormalization factors for the Lagrangian eq.~\eqref{eq:L2} for $a_1=-1/2$, $a_2=-1$, $a_3=1$ and the field eq.~\eqref{27a} are:
\begin{align}
Z_{\bphi} &=1 +
\begin{bmatrix} 1 \\ \overline C_4 \\ \overline C_6 \\ D_4 \\ D_2
\end{bmatrix}^\intercal
 \begin{bmatrix}
 0 & -\frac{1}{2} (n+2) \bar{\lambda}^2 & 0 \\
 2 n \overline{m}^2 & -(n+2) \bar{\lambda} \overline{m}^2 & 2 (n+1) (n+2) \bar{\lambda} \overline{m}^2 \\
 0 & 0 & 0 \\
 (n+2) \overline{m}^2 & -\frac72 (n+2) \bar{\lambda} \overline{m}^2 & (n+2) (n+5) \bar{\lambda} \overline{m}^2 \\
 0 & -2 (n+2) \bar{\lambda}^2 \overline{m}^2 & 0
\end{bmatrix}
 \begin{bmatrix} \left\{ \frac{1}{\epsilon} \right\}_1 \\ \left\{ \frac{1}{\epsilon} \right\}_2 \\ \left\{ \frac{1}{\epsilon^2} \right\}_2 \end{bmatrix} 
 \label{eq:Zphi_phys}
\end{align}
\begin{align}
Z_{\overline m^2} &=1 +
\begin{bmatrix} 1  \\ \overline C_4 \\ \overline C_6 \\ D_4 \\ D_2
\end{bmatrix}^\intercal
\begin{bmatrix}
 (n+2) \bar{\lambda} & -\frac52 (n+2) \bar{\lambda}^2   & (n+2) (n+5) \bar{\lambda}^2 \\
 -4 n \overline{m}^2 & \frac{20}{3} (n+2) \bar{\lambda} \overline{m}^2 & -2  (n+2)(7n+6) \bar{\lambda} \overline{m}^2
  \\
 0 & 0 & -6 (n+2)(n+4) \overline{m}^2  \\
 0 & 0 & 0 \\
 0 & 0 & 0
\end{bmatrix}
 \begin{bmatrix} \left\{ \frac{1}{\epsilon} \right\}_1 \\ \left\{ \frac{1}{\epsilon} \right\}_2 \\ \left\{ \frac{1}{\epsilon^2} \right\}_2 \end{bmatrix} 
\end{align}
\begin{align}
Z_{\bar{\lambda}}\bar{\lambda} &= \bar{\lambda} +
\begin{bmatrix} 1 \\ \overline C_4 \\ \overline C_6 \\ D_4 \\ D_2
\end{bmatrix}^\intercal
\begin{bmatrix}
 (n+8) \bar{\lambda}^2 & -3 (3 n+14) \bar{\lambda}^3 & (n+8)^2 \bar{\lambda}^3 \\
 -8 (n+3) \bar{\lambda} \overline{m}^2 & \frac{8}{3} (22 n+113) \bar{\lambda}^2 \overline{m}^2 & -12 \left(2 n^2+21 n+50\right) \bar{\lambda}^2 \overline{m}^2 \\
 -12 (n+4) \overline{m}^2 & 120 (n+4) \bar{\lambda} \overline{m}^2 & -36 (n+4)(n+10) \bar{\lambda} \overline{m}^2  \\
 0 & 0 & 0 \\
 0 & 0 & 0 
\end{bmatrix}
 \begin{bmatrix} \left\{ \frac{1}{\epsilon} \right\}_1 \\ \left\{ \frac{1}{\epsilon} \right\}_2 \\ \left\{ \frac{1}{\epsilon^2} \right\}_2 \end{bmatrix} 
\end{align}
\begin{align}
Z_{\overline C_4}\overline C_4 &= \overline C_4 +
\begin{bmatrix} 1 \\ \overline C_4 \\ \overline C_6 \\ D_4 \\ D_2
\end{bmatrix}^\intercal
\begin{bmatrix}
 0 & 0 & 0 \\
 2 (n+2) \bar{\lambda} & -\frac{17}{2} (n+2) \bar{\lambda}^2 & 3 (n+2) (n+4) \bar{\lambda}^2 \\
 0 & 0 & 0 \\
 0 & 0 & 0 \\
 0 & 0 & 0 
 \end{bmatrix}
 \begin{bmatrix} \left\{ \frac{1}{\epsilon} \right\}_1 \\ \left\{ \frac{1}{\epsilon} \right\}_2 \\ \left\{ \frac{1}{\epsilon^2} \right\}_2 \end{bmatrix} 
\end{align}
\begin{align}
Z_{\overline C_6}\overline C_6 &= \overline C_6 +
\begin{bmatrix} 1 \\ \overline C_4 \\ \overline C_6 \\ D_4 \\ D_2
\end{bmatrix}^\intercal
\begin{bmatrix}
 0 & 0 & 0 \\
 10 \bar{\lambda}^2 & -\frac23 (23 n+259) \bar{\lambda}^3 & 5 (7 n+62) \bar{\lambda}^3 \\
 3 (n+14) \bar{\lambda} & -\frac{21}{2} (7 n+54) \bar{\lambda}^2 & 3 (n+14)(2n+25) \bar{\lambda}^2
  \\
 0 & 0 & 0 \\
 0 & 0 & 0 
\end{bmatrix}
 \begin{bmatrix} \left\{ \frac{1}{\epsilon} \right\}_1 \\ \left\{ \frac{1}{\epsilon} \right\}_2 \\ \left\{ \frac{1}{\epsilon^2} \right\}_2 \end{bmatrix} 
\end{align}

The additional field redefintion to $\check{\phi}$ in eq.~\eqref{eq:infinitefieldredef} leaves the coupling renormalization factors unchanged, but changes the field renormalization factor to 
\begin{align}
Z_{\check\phi} &=1 +
\begin{bmatrix} 1 \\ \overline C_4 \\ \overline C_6 \\ D_4 \\ D_2
\end{bmatrix}^\intercal
 \begin{bmatrix}
 0 & -\frac{1}{2} (n+2) \bar{\lambda}^2 & 0 \\
 2 n \overline{m}^2 & -(n+2) \bar{\lambda} \overline{m}^2 & 2 (n+1) (n+2) \bar{\lambda} \overline{m}^2 \\
 0 & 0 & 0 \\
 0 & 0 & 0 \\
 0 & 0 & 0 
\end{bmatrix}
 \begin{bmatrix} \left\{ \frac{1}{\epsilon} \right\}_1 \\ \left\{ \frac{1}{\epsilon} \right\}_2 \\ \left\{ \frac{1}{\epsilon^2} \right\}_2 \end{bmatrix} 
\label{eq:Zphi_phys2}
\end{align}

\subsection{\texorpdfstring{$\beta$}{beta}-Functions and \texorpdfstring{$\gamma_\phi$}{gamma\_phi}}\label{sec:physanomdim}

The $\beta$-functions computed from the counterterms in the physical basis are

\begin{align} \nonumber 
	\beta_{\overline m^2} = ~&
\Big\{ 2 (n+2) \bar{\lambda} \overline{m}^2-8 n  \overline{m}^4\overline{C}_4
\Big\}_1+ \Big\{ -10 (n+2) \bar{\lambda}^2 \overline{m}^2+ \frac{80}{3} (n+2)  \bar{\lambda} \overline{m}^4\overline{C}_4
\Big\}_2
\\\nonumber
	\beta_{\bar{\lambda}} = ~&
\Big\{ 2 (n+8) \bar{\lambda}^2 -16 (n+3) \bar{\lambda} \overline{m}^2\overline{C}_4 -24 (n+4)  \overline{m}^2 \overline{C}_6 
\Big\}_1
\\\nonumber
&+ \Big\{ -12 (3 n+14) \bar{\lambda}^3  + \frac{32}{3} (22 n+113)  \bar{\lambda}^2 \overline{m}^2\overline{C}_4 +480 (n+4) \bar{\lambda } \overline{m}^2 \overline{C}_6
\Big\}_2 
\\\nonumber
	\beta_{\overline C_4} = ~&
\Big\{4   (n+2)  \bar{\lambda }\overline{C}_4
\Big\}_1
+ \Big\{ -34 (n+2)  \bar{\lambda }^2\overline{C}_4
\Big\}_2 
\\
	\beta_{\overline C_6} = ~&
\Big\{  20  \bar{\lambda}^2\overline{C}_4 + 6 (n+14) \bar{\lambda} \overline{C}_6
\Big\}_1 
- \Big\{ \frac{8}{3}  (23 n+259)\bar{\lambda}^3 \overline{C}_4 + 42 (7 n+54)  \bar{\lambda}^2\overline{C}_6 
\Big\}_2 
\end{align}
The field anomalous dimension $\gamma_{\bphi}$ from eq.~\eqref{27a} and eq.~\eqref{eq:gammaphidef} including the derivatives w.r.t.\ $D_4$ and $D_2$ is
\begin{align} 
\begin{aligned}
\gamma_{\bphi} = ~&
\Big\{  -2 n  \overline{m}^2\overline{C}_4- (n+2) \overline{m}^2 D_4
\Big\}_1\\ 
&+ \Big\{ (n+2) \bar{\lambda}^2+ 2 (n+2)  \bar{\lambda} \overline{m}^2\overline{C}_4+7 (n+2) \bar{\lambda} \overline{m}^2 D_4+4 (n+2) \bar{\lambda}^2 \overline{m}^2 D_2
\Big\}_2
\end{aligned}
\label{eq:finitegammaphi}.
\end{align}
The field anomalous dimension of $\check{\phi}$ in eq.~\eqref{eq:infinitefieldredef}
is
\begin{align} 
\begin{aligned}
	\gamma_{\check \phi} = ~
\Big\{  -2 n  \overline{m}^2\overline{C}_4
\Big\}_1
&+ \Big\{ (n+2) \bar{\lambda}^2 +2 (n+2) \bar{\lambda} \overline{m}^2\overline{C}_4 \Big\}_2 + \frac{1}{\epsilon} \Big\{ 2 \left(n^2-4\right) \bar{\lambda} \overline{m}^2 \overline{C}_4 \Big\}_2
\end{aligned}
\label{b.no}
\end{align}
and is infinite. This is the same result as computing $\gamma_{\bphi}$ using eq.~\eqref{eq:gammaphidef} but omitting the $\CR$ term.

\end{appendix}

\bibliographystyle{JHEP}
\bibliography{refs.bib}

\providecommand{\href}[2]{#2}\begingroup\raggedright\begin{thebibliography}{10}

\bibitem{Chisholm:1961tha}
J.~S.~R. Chisholm, \emph{{Change of variables in quantum field theories}},
  \href{https://doi.org/10.1016/0029-5582(61)90106-7}{\emph{Nucl. Phys.}
  {\bfseries 26} (1961) 469}.

\bibitem{Politzer:1980me}
H.~D. Politzer, \emph{{Power Corrections at Short Distances}},
  \href{https://doi.org/10.1016/0550-3213(80)90172-8}{\emph{Nucl. Phys. B}
  {\bfseries 172} (1980) 349}.

\bibitem{Arzt:1993gz}
C.~Arzt, \emph{{Reduced effective Lagrangians}},
  \href{https://doi.org/10.1016/0370-2693(94)01419-D}{\emph{Phys. Lett. B}
  {\bfseries 342} (1995) 189}
  [\href{https://arxiv.org/abs/hep-ph/9304230}{{\ttfamily hep-ph/9304230}}].

\bibitem{Manohar:2018aog}
A.~V. Manohar, \emph{{Introduction to Effective Field Theories}},
  \href{https://arxiv.org/abs/1804.05863}{{\ttfamily 1804.05863}}.

\bibitem{Jenkins:2023bls}
E.~E. Jenkins, A.~V. Manohar, L.~Naterop and J.~Pag\`es, \emph{{Two Loop
  Renormalization of Scalar Theories using a Geometric Approach}},
  \href{https://arxiv.org/abs/2310.19883}{{\ttfamily 2310.19883}}.

\bibitem{Bednyakov:2014pia}
A.~V. Bednyakov, A.~F. Pikelner and V.~N. Velizhanin, \emph{{Three-loop SM
  beta-functions for matrix Yukawa couplings}},
  \href{https://doi.org/10.1016/j.physletb.2014.08.049}{\emph{Phys. Lett. B}
  {\bfseries 737} (2014) 129}
  [\href{https://arxiv.org/abs/1406.7171}{{\ttfamily 1406.7171}}].

\bibitem{Herren:2017uxn}
F.~Herren, L.~Mihaila and M.~Steinhauser, \emph{{Gauge and Yukawa coupling beta
  functions of two-Higgs-doublet models to three-loop order}},
  \href{https://doi.org/10.1103/PhysRevD.97.015016}{\emph{Phys. Rev. D}
  {\bfseries 97} (2018) 015016}
  [\href{https://arxiv.org/abs/1712.06614}{{\ttfamily 1712.06614}}], [Erratum:
  Phys.Rev.D 101, 079903 (2020)].

\bibitem{Herren:2021yur}
F.~Herren and A.~E. Thomsen, \emph{{On ambiguities and divergences in
  perturbative renormalization group functions}},
  \href{https://doi.org/10.1007/JHEP06(2021)116}{\emph{JHEP} {\bfseries 06}
  (2021) 116} [\href{https://arxiv.org/abs/2104.07037}{{\ttfamily
  2104.07037}}].

\bibitem{tHooft:1973mfk}
G.~'t~Hooft, \emph{{Dimensional regularization and the renormalization group}},
  \href{https://doi.org/10.1016/0550-3213(73)90376-3}{\emph{Nucl. Phys. B}
  {\bfseries 61} (1973) 455}.

\bibitem{Jenkins:2023rtg}
E.~E. Jenkins, A.~V. Manohar, L.~Naterop and J.~Pag\`es, \emph{{An Algebraic
  Formula for Two Loop Renormalization of Scalar Quantum Field Theory}},
  \href{https://arxiv.org/abs/2308.06315}{{\ttfamily 2308.06315}}.

\bibitem{Herzog:2017bjx}
F.~Herzog and B.~Ruijl, \emph{{The R$^{*}$-operation for Feynman graphs with
  generic numerators}},
  \href{https://doi.org/10.1007/JHEP05(2017)037}{\emph{JHEP} {\bfseries 05}
  (2017) 037} [\href{https://arxiv.org/abs/1703.03776}{{\ttfamily
  1703.03776}}].

\bibitem{Cao:2021cdt}
W.~Cao, F.~Herzog, T.~Melia and J.~R. Nepveu, \emph{{Renormalization and
  non-renormalization of scalar EFTs at higher orders}},
  \href{https://doi.org/10.1007/JHEP09(2021)014}{\emph{JHEP} {\bfseries 09}
  (2021) 014} [\href{https://arxiv.org/abs/2105.12742}{{\ttfamily
  2105.12742}}].

\end{thebibliography}\endgroup

\end{document}